\documentclass[showpacs,floatfix,amsmath,superscriptaddress,amssymb,prb,twocolumn]{revtex4}
\usepackage{latexsym}
\usepackage{graphicx}

\newcommand{\be}{\begin{equation}}
\newcommand{\ee}{\end{equation}}
\newcommand{\ba}{\begin{eqnarray}}
\newcommand{\ea}{\end{eqnarray}}
\newcommand{\baa}{\begin{eqnarray*}}
\newcommand{\eaa}{\end{eqnarray*}}

\begin{document}
\title{Quasiparticle scattering and local density of
states in graphite}
\author{Cristina Bena}
\affiliation{Department of Physics, University of California,
Santa Barbara, CA 93106, USA}
\author{Steven A. Kivelson}
\affiliation{Department of Physics and Astronomy, University of
California, Los Angeles, California 90095-1547, USA}

\date{\today}

\begin{abstract}
We determine the effect of quasiparticle interference on the
spatial variations of the local density of states (LDOS) in
graphite in the neighborhood of an isolated impurity. A number of
characteristic behaviors of interference are identified in the
Fourier transformed spectrum. A comparison between our results and
scanning tunneling microscopy (STM) experiments could provide a
critical test of the range (of energy) of applicability of the
Fermi liquid description of graphite, where some evidence of the
breakdown of Fermi liquid theory has recently been discussed.
Moreover, given the similarity between the band structures of
graphite and  that of nodal quasiparticles in a d-wave
superconductor, a comparison between results in the two materials
is useful for understanding the physics of the cuprates.
\end{abstract}
\maketitle

There has been considerable interest recently in Fourier transform
scanning tunneling spectroscopy (FT-STS) measurements, which may
reveal some rather interesting electronic properties of conducting
materials. In particular, recent experiments performed on certain
cuprate high $T_c$ superconductors \cite{stmscexp,stm1,stm2,stm3,
stmpgexp,vortex,sdavis}, promise to shine some light on the low
energy electronic excitations of the superconducting state, and on
the pseudogap.

In this paper we study a simple two dimensional system, graphite,
which is generally believed to be a Fermi liquid. (Note, however,
that this belief has been questioned in Refs.~\onlinecite{grnfl}.)
Graphite has many similarities to the cuprates; its band structure
consists of two bands  that touch only at the six corners of the
Brillouin zone (BZ). As a result, graphite has nodal linear
dispersing quasiparticle excitations - analogous to those of the
Dirac equation. This makes it in some ways analogous to a d-wave
superconductor, where the nodal quasi particles are expected to
have a (highly anisotropic) Dirac spectrum.  Thus, our interest is
in obtaining an experimentally testable understanding of the
implications of quasiparticle interference, both as a test of the
range of validity of Fermi liquid theory in graphite, and to
provide a fiduciary point for discussions of the more complex
situation in the cuprates.

\begin{figure*}[h]
\begin{center}
\includegraphics[width=7in]{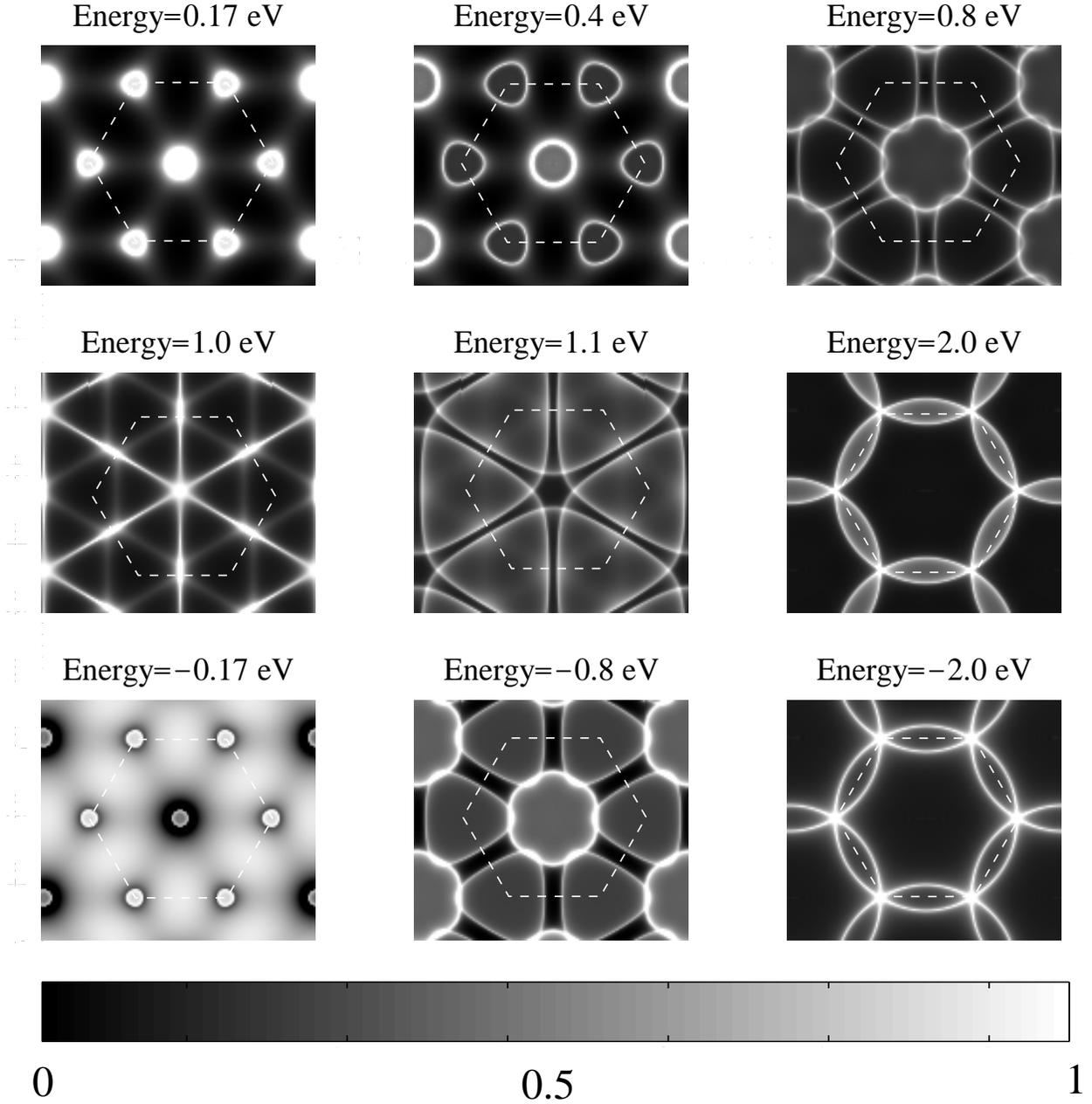}
\end{center}
\vspace{0.15in} \caption{The FT-STS quasiparticle interference
maps in graphite in  the
presence of a single impurity. We plot
$-\rho(q,\omega)$ at the
energies $\omega= 0.17$,
$0.4$, $0.8$, $1$, $1.1$, $2$, $-0.17$, $-0.8$ and $-1.1 {\rm eV}$ .
The $x$ and $y$ axis of each plot
are $q_x$ and $q_y$, and we display our results for
$-1.15 \pi \le q_x \le 1.15 \pi$ and $-\pi \le q_y \le \pi$.
For clarity we draw the BZ in dashed lines.
Note the change in the contours topology with increasing energy.
The plots are drawn using a normalized
linear gray scale, with white being the highest ($1$) intensity,
and black being
the lowest ($0$) intensity, as indicated. The actual values of the FT-LDOS
corresponding
to the lowest ($0$) and highest ($1$) intensity
are different for each energy,
and we find them to be ($-0.12$, $1.73$) for $0.17{\rm eV}$,
($-0.04$, $2.16$) for  $0.4 {\rm eV}$,
($0.27$, $3.6$) for $0.8 {\rm eV}$,
($0.45$, $14.3$) for $1.0 {\rm eV}$,
($0.41$, $3.96$) for $1.1 {\rm eV}$,
($0.22$, $6.45$) for $2.0 {\rm eV}$,
($-8.2$, $-3.26$) for $-0.17 {\rm eV}$,
($-1.1$, $4.07$) for $-0.8 {\rm eV}$,
and ($0.29$, $5.22$) for $-2.0 {\rm eV}$
in arbitrary units}
\label{res}
\end{figure*}

We use a $T$-matrix approach \cite{tmatrix,dhl,ddw} to obtain the
quasiparticle interference spectra for various energies, in the
presence of a single impurity. Our predictions can be easily
tested experimentally, as very clean graphite samples are
relatively easy to obtain. The resulting FT-STS maps can be quite
complex, and contain regions of high intensity. Depending on the
energy, these regions can be circular, triangular, or hexagonal,
as shown in Fig.~\ref{res}. For example, at low energy, the
dominant features in the FT-STS maps are circles centered about
the corners and the center of the BZ. The contours evolve with
increasing energy: their radii increase, the circles centered
about the corners become triangular, while the circle in the
center becomes a hexagon. Beyond the critical energy defined by
the Van Hove singularity, the topology of the contours changes,
and the LDOS exhibits hexagonally shaped lines of high intensity
centered about the center of the BZ. With increasing energy these
lines become smaller
 and circular again, and they disappear altogether at even higher energies.

We now review some general
properties of graphite and explain our $T$-matrix calculation of the FT-STS maps
in
graphite in the presence of a single impurity.

\begin{figure}[h]
\begin{center}
\includegraphics[width=2.5in]{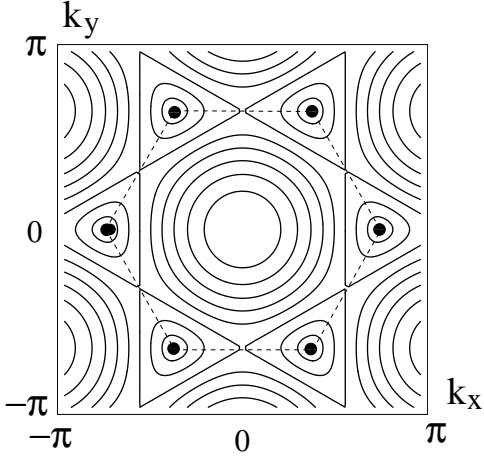}
\end{center}
\vspace{0.15in} \caption{The equal energy contours
$|\epsilon_1(k_x,k_y)|= |\epsilon_2(k_x,k_y)| $ in graphite. The
six points marked by dots correspond to the corners of the BZ
(also the points with $\omega=0$), and the BZ is indicated by the
dashed lines. The band structure at small energies consists of
circular energy contours centered about the corners and the center
of the BZ. At larger energies, the circles change shape and size,
and above some energy $\omega=1 {\rm eV}$ they become hexagons and
then circles centered about the center of the BZ, to disappear
altogether for energies larger than $\omega =3 {\rm eV}$.}
\label{bs}
\end{figure}

Graphite has a hexagonal lattice with one valence electron per
atom and two electrons per unit cell.
% (see Fig. \ref{gr}).
The band structure of graphite  consists of two bands with an
energy dispersion approximated\cite{graphite}  by the simple
particle-hole symmetric tight-binding band structure:

\ba &&\epsilon_{\pm}(k)=
\\&&
=\pm \gamma \sqrt{3+2 \cos(\sqrt{3} k_x a)+ 4
\cos\big(\frac{\sqrt{3}k_x a}{2}\big) \cos\big(\frac{3 k_y
a}{2}\big)},
\nonumber
\ea
where $a$ is the distance between
nearest neighbors,
%(see Fig. \ref{gr})
and $\gamma \approx 1 {\rm eV}$. For simplicity we will set
$\gamma=1 {\rm eV}$ and $a=1$ in the rest of the calculation. The
two bands touch at the corners of the BZ, at points $K_0=2
\pi/9(\pm 2\sqrt{3},0)$, $2 \pi/9(\pm \sqrt{3},\pm 3)$. The band
structure for graphite is depicted in Fig.~\ref{bs}. At low
energies, for wavevectors close to the corners of the BZ, the
dispersion of the nodal quasiparticle excitations is linear
$\epsilon_{\pm}(k)=\pm 3 \gamma |k-K_0|/2$.

\begin{figure}[h]
\begin{center}
\includegraphics[width=2.5in]{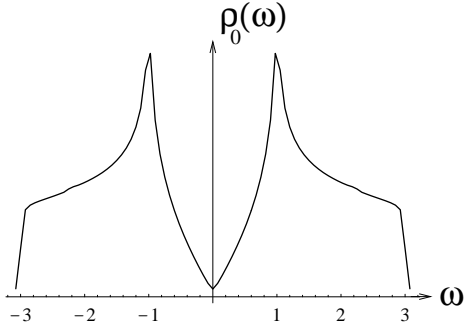}
\end{center}
\vspace{0.15in} \caption{The single particle density of states
$\rho(\omega)$ plotted as a function of energy $\omega$. We note
the v-shaped DOS of small energy and the presence of the van-Hove
singularities at energies of $\pm 1 {\rm eV}$. At energies
$|\omega| \ge 3 {\rm eV}$ the DOS goes to zero. } \label{dos}
\end{figure}

The single particle density of states (DOS) of graphite can be easily obtained by
integrating
the spectral function, $\rho_0(\omega)=\int d^2 k/(4 \pi^2) A(k,\omega)$, where the integral
is
performed over the BZ, and
\ba
A(k,\omega)=-2\sum_{j=1,2}{\rm Im}\left\{\left[\omega-\epsilon_j(k)+i
\delta\right]^{-1}\right\}.
\ea
In the numerical evaluation of the
single particle density of states we took a finite quasiparticle
 inverse lifetime $\delta=0.02 {\rm eV}$.
The result is sketched in Fig.~\ref{dos}. The tiny DOS at $\omega
= 0 $, as well as the rounding of the van Hove singularities, are
due to the finite quasiparticle lifetime. When the quasiparticle
inverse lifetime $\delta$ approaches zero, the density of states
at the Van Hove singularity diverges $\rho_0(\omega_{VH}) \propto
\ln(1/\delta)$, and the density of states at zero energy  goes to
zero, $\rho_0(0) \propto \delta \ln(1/\delta)$.

We note that the density of states is symmetric about the origin
$\omega=0$ and is v-shaped for energies $|\omega|<1 {\rm eV}$. At
the points $\omega_{VH}=\pm 1{\rm eV}$, the density of states
shows van Hove singularities. This is the point where equal energy
contours change topology in the band structure (see
Fig.~\ref{bs}). As expected, the DOS goes to zero at energies
 $|\omega| \ge \pm 3 {\rm eV}$.

Following Refs.~\onlinecite{dhl} and \onlinecite{ddw} we compute
the effect of a single impurity on the LDOS in graphite. We define
a finite temperature (imaginary time) Green's function, \ba
G(k_1,k_2,\tau)=-{\rm Tr}\, e^{-\beta(K-\Omega)}\, {\rm T}_{\tau}
\, \psi_{k_1}(\tau)\psi^+_{k_2}(0), \ea where $K=H-\mu N$,
$e^{-\beta\Omega}={ \rm Tr}\, e^{-\beta K}$, and $ {\rm T}_{\tau}
$ is the imaginary time ordering operator. The impurity scattering
problem can be solved by computing the Fourier transform of the
Green's function from the $T$-matrix formulation \cite{dhl,ddw}:
\ba G(k_1,k_2,i\omega_n)
=G_0(k_1,i\omega_n)T(k_1,k_2,i\omega_n)G_0(k_2,i\omega_n), \ea
where
\begin{eqnarray}
G_0(k,i \omega_n)=\left( \begin{array}{cc}
[i \omega_n-\epsilon_1(k)]^{-1} &0 \\
0 &[i \omega_n-\epsilon_2(k)]^{-1}  \\
\end{array} \right ),
\end{eqnarray}
and \ba &&T(k_1,k_2,i\omega_n)= V(k_1,k_2) \nonumber \\&&
+\sum_{k'}V(k_1,k')G_0(k',i\omega_n)T(k',k_2,i\omega_n). \label{t}
\ea

We will assume that the impurity scattering potential is very
close to a delta function so that $V$ is independent of $k$ and
$k'$,
\begin{eqnarray}
V=v\left( \begin{array}{cc}
1 &1 \\
1 & 1\\
\end{array} \right ).
\end{eqnarray}
For this case we can solve Eq. (\ref{t}), and obtain
\begin{eqnarray}
T(i\omega_n)&=& [I-V {\it{\bar{a}}} \frac{d^2 k}{4 \pi^2}
G_0(k,i\omega_n)]^{-1}V
\\ &=&
\frac{V}{1-v {\it{\bar{a}}}\int \frac{d^2 k}{4 \pi^2}
[G_0^{11}(k,i\omega_n)+G_0^{22}(k,i\omega_n)]}, \nonumber
\end{eqnarray}
where $I$ is the $2 \times 2$ identity matrix, ${\it{\bar{a}}}=3
\sqrt{3}/2$ is the area of the real space unit cell of the
honeycomb lattice, and the integral over $k$ is performed over the
entire Brillouin zone. In the neighborhood of the impurity,
spatial oscillations of the local density of states  are induced.
It is easier to extract information about the quasiparticle
characteristics from the Fourier transform of the LDOS:
\begin{eqnarray}
\rho(q,\omega)=i \sum_{k}g(k,q,\omega), \label{g2}
\end{eqnarray}
where $g(k,q,\omega) \equiv
\sum_{i=1,2}G_{ii}(k,k+q,\omega)-G_{ii}^*(k+q,k,\omega)$, and
$G(k,k+q,\omega)$ is obtained by analytical continuation $i
\omega_n \rightarrow \omega+i\delta$ from $G(k,k+q, i\omega_n)$.
We obtain
 \ba
& &\rho(q,\omega) =-2 \int \frac{d^2 k}{4 \pi^2}  \\
& & \sum_{i=1,2}{\rm Im}\left\{{T(\omega) \over[\omega
-\epsilon_i(k)+ i \delta] [\omega -\epsilon_i(k+q)+ i
\delta]}\right\}, \nonumber \label{q} \ea where $T(\omega)=
 v/\{1-v {\it{\bar{a}}}\int d^2 k/(4 \pi^2) [G_0^{11}(k,\omega+i \delta)+
G_0^{22}(k,\omega+i \delta)]\}$. Eq.(\ref{q}) is analyzed
numerically and the resulting LDOS is plotted in Fig.~\ref{res}.
For our numerical analysis we pick values of $v=2 {\rm eV}$ and
$\delta=0.02 {\rm eV}$. Also we perform our numerical integration
on a $288 \times 240$ grid. We plot our results on a $144 \times
120$ grid.  We have confirmed that the results have converged in
the sense that changing the size of the grid leads to no
observable changes in the results.

At low energies, our analysis reveals that the dominant features
in the LDOS spectra are circular contours of radius $4 \omega/3
\gamma$, centered about the corners and the center of the BZ. The
shape of the contours changes with energy, and their size
increases. At the energy $\omega= \omega_{VH}=1 {\rm eV}$, the
spectral features change to hexagonal shaped contours centered
about the center of the BZ. With increasing energy the contours
become smaller and circular, and disappear for energies above
$\omega=3 {\rm eV}$.

A few comments about our results are in order. We note that the
intensity inside and outside the circles is not zero but
finite\cite{RMP}, which, depending on the quasiparticle inverse
lifetime $\delta$, can blur the interiors of the circles,
especially at low energies.

\begin{figure}[h]
\begin{center}
\includegraphics[width=3in]{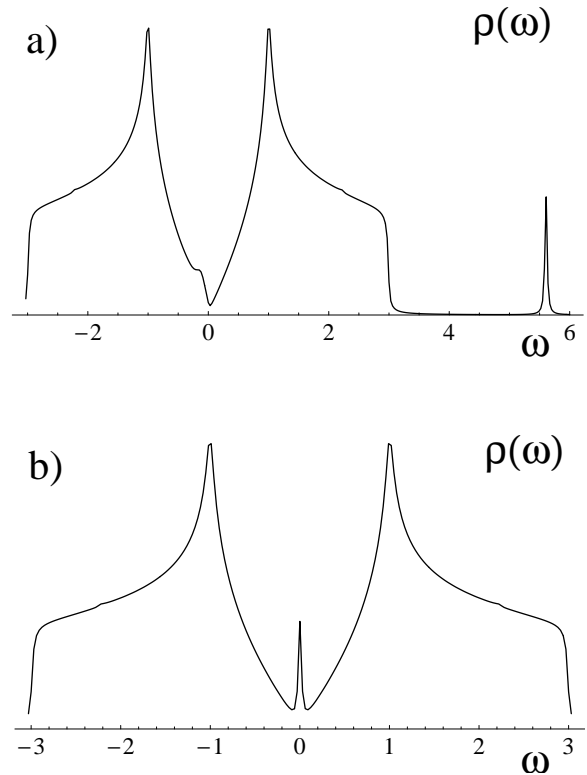}
\end{center}
\vspace{0.15in} \caption{The spatially averaged single particle
density of states $\rho(\omega)$, in the presence of an impurity
of strength a) $v=2.5 {\rm eV}$, and b) $v=\infty$.} \label{vdep}
\end{figure}

The topology of the contours is identical for negative and
positive energies.  However the weight, and in some cases the sign
of the peaks relative to the background, as well as the average
spectral intensities are different, in particular at low energies.
The asymmetry of the FT-STS spectra with respect to energy is not
a direct  band structure effect, since for graphite the band
structure is particle-hole symmetric. It is an effect of the
impurity scattering. We will discuss this asymmetry also in
connection to the effect of the impurity scattering on the average
density of states in Fig.~\ref{vdep}.

For positive energies,
the results presented here do not depend sensitively on the strength
of the impurity
potential, $v$. We have repeated our
calculation for the FT-STS spectra for
different values of $v$, including the limit of infinite scattering
 strength (``unitarity limit'') and
the changes are not qualitatively significant. However, at
negative energies, our FT-STS pictures evolve with increasing $v$,
and at large scattering strengths they become similar to their
positive energy counterparts.

Fermi liquid theory is most robust at small probing energies, and
at these energies we would expect the experimentally observed
features to have the best agreement with the theory. As noted in
Fig.~\ref{res}, at small positive energies the amplitude of the
LDOS oscillations due to the impurity scattering is lower than at
high energies, which may make the experimental observation of
these oscillations more difficult. However, the fractional
amplitude of the LDOS oscillations at small negative energies is
larger than at higher energies.

The Van-Hove singularities which appear at high energies, are a
``big" feature, so survive in some form unless there is a very
serious failure of the quasiparticle picture.  However, it is
still possible that a strongly energy dependent quasiparticle
lifetime could considerably mute the Van Hove singularities, while
leaving the Fermi liquid behavior intact at much lower energies.

We note that at low energy the position of the dots/circles is
determined by symmetry, while at higher energies the shape and the
position of the high intensity lines is sensitive to the assumed
band structure. Another interesting point is that in a two
dimensional system, the general structure of singularities is
lines instead of points\cite{RMP}. This yields  lines of high
intensity regions in the in FT-STS spectra due to quasiparticle
interference, as opposed to points\cite{RMP}.

Another interesting aspect we analyzed is how the spatially
averaged single particle density of states
$\rho(\omega)=\rho_0(\omega)+ 2 c_{imp}\rho(q=0,\omega)$ changes
with the scattering strength $v$. Here $c_{imp}$ is the
concentration of impurities. We assume that $c_{imp}$ is small, so
the effects of the impurities are uncorrelated. The resulting
density of states depends on the value of the impurity potential.
For comparison, in Fig.~\ref{vdep}, we give $\rho(\omega)$ for
$v=2.5 {\rm eV}$ (a), and for $v=\infty$ (b). We fix the impurity
concentration to $c=1\%$. As expected, in the unitarity limit the
effect of the impurity is to create a sharp peak centered at zero
energy. For smaller impurity potential, the effect of the impurity
is to create a smaller and broader peak at negative energy. One
also notes the existence of a sharp anti-bound state at an energy
larger than $3 {\rm eV}$. The position of the anti-bound state
moves to larger and larger energies as one increases the
scattering strength.

We also performed an analytical calculation of the local density
of states as a function of position at low energy, where we can
approximate the spectrum by a linear (Dirac) dispersion
$\omega=\pm v_F |k|$.  We focused on the case of a single impurity
scattering. For points far from the impurity we found the
correction to the density of states due to impurity scattering to
have the form:

\begin{equation}
\rho(\vec{x},\omega)=\sum_{i=0}^{m} \frac{A_i}{|\vec{x}|}
\cos\Big(\frac{2\omega }{v_F} |\vec{x}| +\vec{Q}_i \cdot
\vec{x}+\phi_i\Big),
\end{equation}
where $v_F=3 \gamma/2$ is the Fermi velocity. Here $m=2$ is the
number of the independent inter-nodal scattering vectors (that
cannot be obtained from other scattering vectors by translation by
a reciprocal lattice vector or by reflection). We can take, for
example, $\vec{Q}_0=(0,0)$, $\vec{Q}_1=2 \pi/9(2\sqrt{3},0)$,
$\vec{Q}_2=2 \pi/9(\sqrt{3},3)$. The coefficients $A_i$ and
$\phi_i$ depend on energy and on the impurity characteristics.  As
expected, at large $|\vec{x}|$, the density of states decays as
$1/|\vec{x}|$. If one takes into account also the finite
quasiparticle lifetime the above equation for the density of
states changes to
\begin{equation}
\rho(\vec{x},\omega)\rightarrow \rho(\vec{x},\omega)e^{-|\vec{x}|
\tau/v_F},
\end{equation}
where $\tau=\tau(\omega)$ is the quasiparticle lifetime. A
measurement of the density of states as a function of position
thus reflects the position of the nodes ($\vec{Q}_i$), the
quasiparticle dispersion, $v_F$, and the energy dependence of the
quasiparticle lifetime.

%To conclude, we have performed a $T$-matrix analysis of the FT-STS
%quasiparticle interference patterns in graphite in the presence of
%a single impurity. Our results show a variety of patterns which
%change with probing energy. The single particle DOS shows van Hove
%singularities at the points at which the patterns change topology.
%It would be interesting to compare both the single particle DOS
%and the FT-STS maps with the experimental results. Also, it would
%be of great importance to compare the experimental and theoretical
%data in graphite with the ones in cuprates, to help elucidating
%some recent puzzles in the FT-STS measurement in cuprates.
%Experiments performed in  cuprates reveal FT-STS patterns in good
%agreement with the quasiparticle interference picture, but also
%reveal new intriguing features which can be a sign of incipient
%order. As it is believed that graphite does not have any incipient
%order, it is our hope that the comparison between theory and
%experiment in graphite and in the cuprates will help distinguish
%between such scenarios.

\acknowledgments We would like to thank Hongwen Jiang and Jianping
Hu for interesting discussions. C. B. has been supported by the
NSF under Grant No. DMR-9985255, and also by funds from the
Broida-Hirschfeller Foundation, the A. P. Sloan Foundation and the
Packard Foundation. S. K. has been supported by the DOE under
Grant No. DE-FG03-00ER45798.

%\begin{widetext}
%\end{widetext}

\end{document}